\documentclass[reprint,aps,prb,superscriptaddress]{revtex4-1}
\usepackage{graphicx}
\usepackage[version=4]{mhchem}
\usepackage[colorlinks, citecolor=blue, linkcolor=blue, urlcolor=blue, breaklinks]{hyperref}
\usepackage{siunitx, xspace}
\DeclareSIUnit[]\muB{\text{\ensuremath{\mu_{\textup{B}}}}}
\bibpunct{[}{]}{,}{n}{}{}

\let\vaccent=\v
\let\v\undefined
\def\v#1{\mathbf{#1}} 
\newcommand{\pd}[2]{\frac{\partial #1}{\partial #2}}

\begin{document}

\title{Maximum entropy formalism for the analytic continuation \\ of matrix-valued Green's functions}

\author{Gernot J.~Kraberger}
\affiliation{Institute of Theoretical and Computational Physics,
Graz University of Technology, NAWI Graz, 8010 Graz, Austria}
\author{Robert Triebl}
\affiliation{Institute of Theoretical and Computational Physics,
Graz University of Technology, NAWI Graz, 8010 Graz, Austria}
\author{Manuel Zingl}
\affiliation{Institute of Theoretical and Computational Physics,
Graz University of Technology, NAWI Graz, 8010 Graz, Austria}
\author{Markus Aichhorn}
\email[]{aichhorn@tugraz.at}
\affiliation{Institute of Theoretical and Computational Physics,
Graz University of Technology, NAWI Graz, 8010 Graz, Austria}

\date{\today}

\begin{abstract}
We present a generalization of the maximum entropy method to the
analytic continuation of matrix-valued Green's functions.
To treat off-diagonal elements correctly based on Bayesian probability theory, the entropy term has to be
extended for spectral functions that are possibly negative in some
frequency ranges.
In that way, all matrix elements of the
Green's function matrix can be analytically continued; we
introduce a computationally cheap element-wise method for this purpose.
However, this method cannot ensure important constraints on the mathematical
properties of the resulting spectral functions, namely positive
semidefiniteness and Hermiticity. To improve on this, we present a
full matrix formalism, where all matrix elements are treated
simultaneously.  
We show the capabilities of these methods using insulating and metallic
dynamical mean-field theory (DMFT) Green's functions as test cases. Finally, we apply the methods to
realistic material calculations for \ce{LaTiO3}, where off-diagonal
matrix elements in the Green's function appear due to the distorted
crystal structure. 
\end{abstract}

\maketitle

\section{Introduction}

In condensed matter physics, response functions are often calculated in imaginary-time formulation, especially when electronic correlations are taken into account.
This is not only true for numerical approaches like quantum Monte Carlo~\cite{hirsch_monte_1986,foulkes_quantum_2001,gull_continuous-time_2011}, but also for perturbative techniques such as the random phase approximation~\cite{harl_accurate_2009, kaltak_cubic_2014, liu_cubic_2016}.
However, these quantities cannot be directly related to measurable quantities in real frequency. 
Quite generally, the Wick rotation $i\tau\to t$, where $\tau$ is the imaginary-time argument and $t$ is the real-time argument (or equivalently $i\omega_n\to\omega$, with the $n$th fermionic Matsubara frequency $\omega_n = (2n+1)\pi/\beta$ and the real frequency $\omega$), transforms the calculated quantities to real frequencies. 
In practice, this analytic continuation (AC) is not possible straightforwardly, since the kernel of this mapping is ill-conditioned when going from imaginary times to real frequencies. As a result of the kernel being ill-conditioned, small changes of the input will correspond to largely different outputs, rendering the inversion of this problem highly unstable due to numerical noise, where even an error at the level of machine precision can lead to nonsensical results in practice.

This fact has lead to the development of a plethora of different methods trying to efficiently perform the AC. 
Among them are series expansions (e.g.,\ the Pad\'e method~\cite{ferris-prabhu_numerical_1973,vidberg_pade, beach_reliable_2000}), information-theoretical approaches such as the maximum entropy method (MEM)~\cite{silver_maximum-entropy_1990,Gubernatis_MEM,gull_skilling_MEM_imageproc, bao_fesom} and stochastic methods~\cite{SOM_Sandvik, SOM_Beach, SOM_Mishchenkov, fuchs_mem_som,sandvik_2016}. 
Other algorithms based on singular value decomposition (SVD)~\cite{creffield_svd}, machine learning~\cite{arsenault_projected_2016} or sparse modeling~\cite{otsuki_sparse} tackling the AC have also been presented. 
Despite all those interesting other developments, the workhorse method for the AC of noisy Monte Carlo data is the MEM.

The methods based on the MEM are well established for the diagonal elements of the Green's function, where the corresponding spectral function can be interpreted as a probability distribution (non-negative normalizable function).
There are several freely available codes performing this task, such as $\Omega$\textsc{maxent}~\cite{PhysRevE.94.023303} and the \textsc{maxent} code by Levy {\em et al.}~\cite{levy_implementation_2016}.

Nowadays, numerical algorithms do also provide imaginary-time solutions for off-diagonal Green's functions, e.g.,\ in the multiorbital DFT~+~DMFT~\cite{DMFT0, DMFT1, DMFT2} context relevant for real-material applications.
However, due to the lack of reliable methods for performing the AC of the whole Green's function matrix, still the off-diagonal elements are often neglected on different levels of the calculation. 
One strategy is to transform the impurity problem to some local basis, where the Hamiltonian and hybridization functions are as diagonal as possible and to neglect the off-diagonal elements in the solution of the impurity problem~\cite{LaTiO3_Dang}. 
However, this is an uncontrolled approximation because it is impossible to check the accuracy of this approximation without actually taking the off-diagonal elements into account.

Particularly important is a proper AC for the self-energy. The full matrix form of the self-energy on the real axis is required in the Dyson equation to calculate lattice ($k$-dependent) quantities of interest. 
Without ensuring analytic properties such as positive semidefiniteness of this matrix, the results for quantities such as the $k$-dependent spectral function $A({\mathbf k}, \omega)$ or derived quantities (e.g., transport, optics) are physically questionable. 
We will present a method to remedy this problem.

For certain cases, the AC of off-diagonal Green's functions has been tackled before:
In general, it is possible to construct an auxiliary Green's function by adding a (possibly frequency-dependent) shift to the off-diagonal ele\-ments of the spectral function so that their positivity is ensured. Then, they can be treated with the MEM~\cite{tomczak_effective_2007, jarrell_maximum_2012, reymbaut_maximum_2015}.
An example of a work where off-diagonal elements of the impurity spectral function are calculated are the DFT+DMFT calculations of the two perovskites \ce{LaVO3} and \ce{YVO3}~\cite{LaVO3_Raychaudhury}.
Additionally, a stochastic regularization method also suitable for off-diagonal elements has been proposed~\cite{krivenko_analytic}.
However, these methods cannot ensure important matrix properties, e.g.,\ positive semidefiniteness and Hermiticity of the spectral function.
Additionally, given the probability theoretical background of the MEM~\cite{skilling_1989}, it is unclear how the shift method fits into this theoretical framework.

In other disciplines, such as astronomy and nuclear magnetic resonance (NMR), the MEM has been successfully extended to extract data without the constraint of non-negativity~\cite{laue_1985, sibisi_1990,maisinger_1997, jones_1998, hobson_1998}. 
This generalization is not straightforward, as non-negative functions cannot be directly interpreted as probability distributions. 

But even with this generalization, important matrix properties are not respected. 
The purpose of this paper is, thus, to introduce a consistent {\em matrix} formulation of the MEM completely from probability theory. 
Using the full matrix enables us to consistently formulate the constraint that the resulting spectral functions are indeed positive semidefinite and Hermitian. 

The paper is organized as follows: First, we present the probability theoretical background of the continuation of matrix-valued Green's functions and some computational and implementation details in Sec.~\ref{sec:methodology}. 
In Sec.~\ref{sec:bethe-model}, we perform a benchmark of the MEM and discuss some practical considerations using a DMFT calculation for a model system. 
Finally, in Sec.~\ref{sec:LaTiO3}, we apply the introduced methodology within the framework of DFT+DMFT to the strongly correlated perovskite \ce{LaTiO3}. 

\section{Methodologys and Theory}
\label{sec:methodology}
\subsection{Basic principles of the maximum entropy method}

The retarded one-electron Green's function $G(\omega+i0^+)$ and the Matsubara Green's function $G(i \omega_n)$ are related through the analyticity of $G(z)$ in the whole complex plane with the exception of the poles below the real axis.
This connection is explicit by writing the Green's function $G(z)$ in terms of the spectral function~$A(\omega)$ as
\begin{equation}
    G_{ab}(z) = \int d\omega \frac{A_{ab}(\omega)}{z - \omega}.
    \label{eq:spectral-representation}
\end{equation}
In general, both $G(z)$ and $A(\omega)$ are matrix-valued (with indices $a$, $b$), but Eq.~\eqref{eq:spectral-representation} is valid for each matrix element separately.
For a given $G_{ab}(\omega+i0^+)$, the matrix-valued $A_{ab}(\omega)$ can be obtained as
\begin{equation}
    A_{ab}(\omega) = \frac{i}{2\pi} \left[G_{ab}(\omega+i0^+) - G_{ba}^*(\omega+i0^+)\right].
\end{equation}
Note that for matrices, the spectral function is not proportional to the element-wise imaginary part of the Green's function.

A drawback of expression~\eqref{eq:spectral-representation} is that the real and imaginary parts of $G$ and $A$ are coupled due to the fact that $z$ is complex-valued.
This is avoided by Fourier-transforming $G(z=i\omega_n)$ to the imaginary time Green's function $G(\tau)$ at inverse temperature $\beta$;
\begin{equation}
 G_{ab}(\tau) = \int d\omega  \; \frac{e^{-\omega\tau}}{1+e^{-\omega\beta}}A_{ab}(\omega).
 \label{eq:Gtau-from-Aw}
\end{equation}
The real part of the spectral function is only connected to the real part of $G(\tau)$, and analogously for the imaginary part.
In the following, we will first recapitulate the maximum entropy theory for a real-valued single-orbital problem as presented in Ref.~\cite{Gubernatis_MEM} and later generalize to matrix-valued problems.

In order to handle this problem numerically, the functions $G(\tau)$ and $A(\omega)$ in Eq.~\eqref{eq:Gtau-from-Aw} can be discretized to vectors $G_n= G(\tau_n)$ and $A_m = A(\omega_m)$; then, Eq.~\eqref{eq:Gtau-from-Aw} can be formulated as
\begin{equation}
\v{G} = K \v{A},
\label{eq:Gtau-from-Aw-matrix}
\end{equation}
where the matrix
\begin{equation}
 K_{nm} = \frac{e^{-\omega_m\tau_n}}{1+e^{-\omega_m\beta}} \Delta \omega_m
\end{equation}
is the kernel of the transformation.
Calculating $G(\tau)$ from $A(\omega)$ is straightforward, but the inversion of the matrix equation~\eqref{eq:Gtau-from-Aw-matrix}, i.e.,\ calculating $\v{A}$ via $\v{A} = K^{-1}\v{G}$, is an \emph{ill-posed} problem.
To be more specific, the condition number of $K$ is very large due to the exponential decay of $K_{nm}$ with $\omega_m$ and $\tau_n$, so that the direct inversion of $K$ is numerically not feasible by standard techniques. 

The task of the AC is to find an approximate spectral function $\v{A}$ whose reconstructed Green's function $\v{G}_{rec} = K\v{A}$ reproduces the main features of the given data $\v{G}$, but does not follow the noise (note that here and in the following we use $\v{G}$ and $\v{A}$ for the numerical quantities to keep the notation simple).
However, a bare minimization of the misfit $\chi^2 (\v{A}) = (K \v{A} - \v{G})^T C^{-1} (K \v{A} - \v{G})$, with the covariance matrix $C$, leads to an uncontrollable error~\cite{beach_reliable_2000}.

One efficient way to regularize this ill-posed problem is to add an entropic term $S(A)$.
This leads to the maximum entropy method (MEM), where one does not minimize $\chi^2 (A)$, but 
\begin{equation}\label{eq:Q}
 Q_\alpha(A) = \frac12 \chi^2 (A) - \alpha S(A).
\end{equation}
The prefactor of the entropy, usually denoted $\alpha$, is a hyperparameter that is introduced \emph{ad hoc} and needs to be specified. 
The way to choose $\alpha$ marks various flavors of the maximum entropy approach and will be discussed later (Sec.~\ref{sec:hyper}).
This regularization with an entropy has been put on a rigorous probabilistic footing by Skilling in 1989, using Bayesian methods~\cite{skilling_1989}.
He showed that the only consistent way to choose the entropy for a non-negative function $A(\omega)$ is
\begin{equation}
    S(A) = \int d\omega \left[A(\omega) - D(\omega) - A(\omega) \log\frac{A(\omega)}{D(\omega)} \right],
    \label{eq:entropy-conventional}
\end{equation}
where $D(\omega)$ is the default model.
The default model influences the result in two ways (see Appendix~\ref{sec:prob} for details): First, it defines the maximum of the prior distribution, which means that in the limit of large $\alpha$ one has $A(\omega) \rightarrow D(\omega)$. Second, it is also related to the width of the distribution, since the variance of the prior distribution is proportional to $D(\omega)$.
Unless otherwise specified (see, especially, Sec.~\ref{sec:poormatrix}), we use a flat $D(\omega)$, corresponding to no prior know\-ledge.

\subsection{Hyperparameter $\alpha$}\label{sec:hyper}
The simplest way to determine $\alpha$ is to choose it such that $\chi^2$ equals the number of $\tau$ points~\cite{gull_1978, gull_skilling_MEM_imageproc}, which is today known as the historic MEM. 
Usually, it tends to underfit the data~\cite{titterington_1985}.
Other, more sophisticated ways are delivered by the probabilistic picture of Skilling and Gull~\cite{skilling_1989,gull_1989}, which are recapitulated in Appendix~\ref{sec:prob}.
Two frequently used flavors are the classical MEM~\cite{gull_1989} and the Bryan MEM~\cite{bryan_1990}.
A disadvantage is that these probabilistic methods tend to overfit the data as the probability is only evaluated approximately in practice (see Appendix~\ref{sec:prob})~\cite{vdl_1999, hohenadler_2005}.
Furthermore, all methods presented so far strongly depend on the provided covariance matrix $C$. 
If the statistical error of Monte Carlo measurements, for example, is not estimated accurately, the data could be over- or underfitted. 

A rather heuristic approach to overcome these problems is not to consider probabilities, but rather the quality of the reconstruction as a function of $\alpha$.
One way to quantify this is to detect the characteristic kink in the function $\log \chi^2(\log \alpha)$, which indicates the boundary between the noise-fitting and information-fitting regimes~\cite{PhysRevE.94.023303}.
In the noise-fitting region, $\log \chi^2(\log \alpha)$ is essentially constant, while in the information-fitting region, it behaves linearly.
In this approach, the optimal $\alpha$ is at the crossover of these regimes, which can be detected, e.g.,\ through the maximum of the second derivative $\partial^2 \log \chi^2 / \partial (\log\alpha)^2$, as implemented the $\Omega$\textsc{maxent} code~\cite{PhysRevE.94.023303}.

We propose another way, which is to fit a piecewise linear function to $\log \chi^2(\log \alpha)$, consisting of two straight lines: one for the noise-fitting region (with slope zero) and one for the information-fitting region.
The intersection of the two lines, and hence the optimal $\alpha$, is determined such that the overall fit residual is minimized.
This way of determining the optimal $\alpha$ is used throughout the rest of this paper, as it turns out to be stable even in difficult cases where the curvature of $\log \chi^2(\log \alpha)$ shows multiple local maxima.

\subsection{Positive-negative MEM}\label{sec:posneg}
For the case of non-matrix-valued or diagonal spectral functions, the MEM described so far became a standard tool used in many different contexts.
However, this ordinary MEM is only rigorous for non-negative, additive functions~\cite{skilling_1989}.
Nonzero off-diagonal elements of spectral functions clearly violate the non-negativity since their norm
\begin{equation}
    \int d\omega A_{ab}(\omega) = \delta_{ab}
\end{equation}
is zero (this follows directly from the Lehmann representation of the spectral
function and the anticommutation relations of fermionic operators).
Keeping the additivity, one could imagine that the off-diagonal spectral functions originate from a subtraction of two artificial positive functions, i.e.,\ $A(\omega) = A^+(\omega) - A^-(\omega)$.
Assuming independence of $A^+(\omega)$ and $A^-(\omega)$, the resulting entropy is the sum of the respective entropies
\begin{equation} \label{eq:entropy-sum}
S(A^+(\omega),A^-(\omega)) = S(A^+(\omega)) + S(A^-(\omega)),
\end{equation}
which was first used for the analysis of NMR spectra~\cite{laue_1985, sibisi_1990}.
To illustrate the plausibility of this entropy, we use the analogy of the horde of monkeys, which has a long tradition in the field of Bayesian methods.
The conventional entropy can be explained by monkeys randomly throwing balls into slots which correspond to different frequencies $\omega_i$ on a grid~\cite{skilling_1989}.
Then, the number of balls in each slot, related to $A(\omega_i)$, obeys a Poisson distribution with a mean value given beforehand, related to the default model $D(\omega_i)$.
From now on, the $\omega_i$~dependence of $A$ and $D$ is dropped for simplicity. 
The subtraction of two positive functions $A= A^+ - A^-$ can be understood with two different hordes of monkeys, one throwing ``positive'' balls and one throwing ``negative'' balls.
Individually, both the number of positive ($\sim A^+$) and negative ($\sim A^-$) balls again obey a Poisson distribution.
The total number of balls in each slot, however, follows a Skellam distribution, which is the convolution of two Poisson distributions.
The entropy $S$ describing this process depends on both $A^+$ and $A^-$ [see Eq.~\eqref{eq:entropy-sum}].
Due to the independence of $A^+$ and $A^-$, $S(A^+)$ and $S(A^-)$ follow the same functional form as the conventional entropy, Eq.~\eqref{eq:entropy-conventional}, that stems from the Poisson distribution.
Thus,
\begin{eqnarray}
 S(A^+,A^-)& = & \int d\omega \Big[A^+ - D^+ - A^+ \log \frac{A^+}{D^+} \label{eq:plus-minus-entropy}\\*
 & &+ A^- - D^- - A^- \log \frac{A^-}{D^-}\Big].\nonumber
\end{eqnarray}
The fact that two default models ($D^+$ and $D^-$) enter will be discussed later.
Several configurations of positive and negative balls give the same net number of balls ($\sim A$), since only their difference $A^+-A^-$ matters. Hence, an additional superfluous degree of freedom is present once two hordes are acting.
Just as the two Poisson distributions of the respective balls lead to a Skellam distribution by integrating out the additional degree of freedom, a reduction of the parameter space from $A^+$ and $A^-$ to $A = A^+-A^-$ leads to an entropy $S^\pm(A)$ that differs from the conventional entropy in Eq.~\eqref{eq:entropy-conventional}.
The derivation of $S^\pm(A)$ was first carried out in the context of cosmic microwave background radiation~\cite{maisinger_1997, jones_1998, hobson_1998}, an available software package providing this entropy is \textsc{memsys5}~\cite{gull_quantified_1999}. 
This framework is recapitulated here in the context of spectral functions.

The main objective of the MEM is to minimize $Q_\alpha$ as given by Eq.~\eqref{eq:Q}, but the entropy $S$ depends now on both $A^+$ and $A^-$ as shown in Eq.~\eqref{eq:plus-minus-entropy}.
The minimum of
\begin{equation}
Q_\alpha(A^+,A^-) = \frac12 \chi^2(A = A^+ - A^-) - \alpha S(A^+,A^-)
\end{equation}
has to be found with respect to both $A^+$ and $A^-$.
The misfit $\chi^2$ only depends on the difference $A = A^+-A^-$. For any fixed $A$, the minimum of $Q_\alpha(A,A^+)$ is, therefore, realized for the particular choice of $A^+$ and $A^-$ that maximizes the entropy under the constraint that $A = A^+-A^-$.
Expressing $A^-$ in terms of $A$ via $A^- = A^+-A$, the minimum of $Q_\alpha(A,A^+)$ with respect to $A^+$ is given by
\begin{eqnarray}
 A^+& = &\frac{\sqrt{A^2+4D^+D^-}+A}{2},\label{eq:A-interal-opt}\\*
 A^-& = &\frac{\sqrt{A^2+4D^+D^-}-A}{2}.
\end{eqnarray}
A new entropy for functions that can be both positive and negative is then obtained by $S^\pm(A) = S\left(A^+(A),A^-(A)\right)$, which we call positive-negative entropy and which reads~\cite{hobson_1998}
\begin{eqnarray}
 S^\pm(A)& = &\int d\omega \Bigg[\sqrt{A^2 + 4D^+D^-} - D^+ - D^- \label{eq:entropy_pm}\\*
 & &- A \log \frac{\sqrt{A^2 + 4D^+D^-} + A}{2 D^+}\Bigg]. \nonumber
\end{eqnarray}
The Bayesian probabilistic interpretation of this entropy is described in Appendix~\ref{sec:prob}.
In the special case $D^-=0$, the limit of purely positive functions is recovered since then $S^\pm(A) = S(A)$ [the latter being the conventional entropy from Eq.~\eqref{eq:entropy-conventional}].

\begin{figure}
    \centering
    \includegraphics[width=0.95\linewidth]{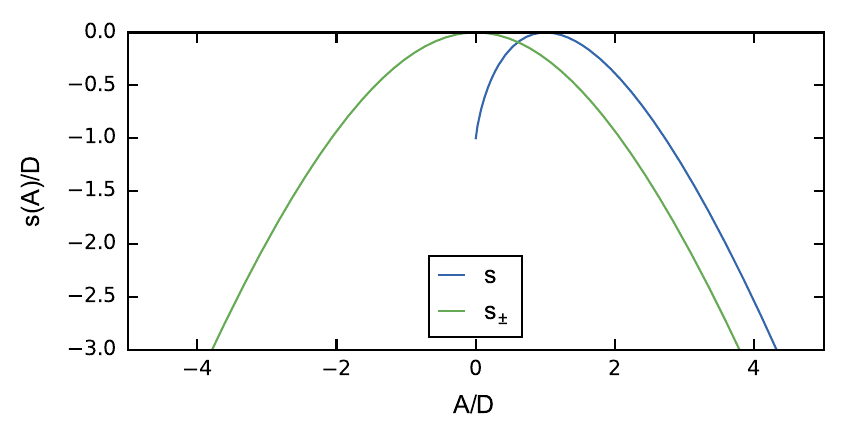}
    \caption{
    Comparison of the ``entropy density'' of non-negative spectral functions $s(A)$ compatible to Eq.~\eqref{eq:entropy-conventional} and of positive and negative spectral functions $s^\pm(A)$ compatible to Eq.~\eqref{eq:entropy_pm}.
    The entropy is related to the entropy density via $S(A) = \int d\omega\ s(A(\omega))$.
    Here, $D^+ = D^- = D$ was assumed.
    }
    \label{fig:entropy-densities}
\end{figure}

Next, we want to compare the conventional entropy, Eq.~\eqref{eq:entropy-conventional}, to the positive-negative entropy, Eq.~\eqref{eq:entropy_pm}.
This comparison can be performed on the level of the integrand of the expression for the entropy, which we refer to as the entropy density $s(A(\omega))$.
At a particular frequency value $\omega_i$, the entropy density depends only on the function value of the spectral function $A(\omega_i)$ and the default model $D(\omega_i)$.
We therefore plot the entropy density $s$ depending on the function value $A$ at any given $\omega_i$ in Fig.~\ref{fig:entropy-densities}.
The ordinary entropy density $s(A)$ (blue line) is just defined for positive $A$.
Within this definition space, it is concave with a maximum at $A = D$. 
The variance of the prior distribution around this maximum is also proportional to $D$ (see Appendix~\ref{sec:prob}).
In the case of $S^\pm$, two default models $D^+$ and $D^-$ are needed, each determining the maximum of the respective spectral functions $A^+$ and $A^-$ as well as the accompanying variances.
Usually, no additional knowledge about $A$ is available and, therefore, one has to choose $D^+ = D^- = D$ for symmetry reasons (green line in Fig.~\ref{fig:entropy-densities}).
This is the case for the off-diagonal elements of the spectral functions studied in this work; however, the general case is discussed in Appendix~\ref{sec:prob}.
For $D^+ = D^- = D$, the maximum of the entropy is at $A=0$, with a prior variance proportional to $D$.
This demonstrates a fundamental difference of the role of the default model in the conventional and in the positive-negative entropy; in the former, the default model defines both the maximum and the variance of the entropy density, while for the latter it only punishes large values of $|A|/D$.
This also means that for $\alpha\to\infty$ the minimization of $Q_\alpha(A)$ gives $A=0$ for the positive-negative entropy, in contrast to $A=D$ for the conventional entropy.

We note in passing that off-diagonal elements of the spectral function can, of course, be complex-valued.
Then, the real and imaginary part of $A(\omega)$ and, correspondingly, $G(\tau)$ can (in principle) be treated separately.
The misfit $\chi^2$ and the entropy $S$ for a complex function can therefore be just summed up to a total $\chi^2$ and $S$.
This straightforward generalization of the method is applicable to this and the following sections, but for simplicity we limit ourselves to real-valued spectral functions.

\subsection{Reduction of the parameter space}\label{sec:reduction}
In Ref.~\cite{bryan_1990}, Bryan presents an algorithm that works in the space of the singular values of $K$, by means of the singular value decomposition (SVD)
\footnote{We would like to point out that the use of SVD in Bryan's algorithm in the MEM is just for the purpose of choosing a convenient basis for finding the minimum of $Q_\alpha$. This is fundamentally different from methods where the SVD with a cutoff is used directly for the AC without a regularization by an entropy term (see, e.g.,\ Ref.~\cite{creffield_svd}) or in conjunction with a $L_1$ regularization to achieve sparseness in the SVD space, as in Ref.~\cite{otsuki_sparse}.}
\begin{equation}
K = U \Xi V^T.
\end{equation}
When the problem is discretized with $N_\tau$ points on the $\tau$~axis and $N_\omega$ points on the $\omega$~axis, the kernel $K$ is a $N_\tau \times N_\omega$ matrix (we use $684\times 200$) which gets decomposed into the singular-vector matrices $U$ of dimension $N_\tau \times N_\Xi$ and $V$ of dimension $N_\omega \times N_\Xi$ as well as the diagonal matrix $\Xi$ of the singular values.
In principle, the number $N_\Xi$ of singular values is given by $\min (N_\tau,N_\omega)$; however, many singular values are on the order of machine precision.
Therefore, in practice all singular values below a small threshold ($10^{-14}$ for the results in this paper) can be discarded.
This also means that the full vector space of $\v{A}$, where $A_i = A(\omega_i)$, is larger than necessary (for our calculations, $N_\omega=200$, while $N_\Xi=45$).

One important ingredient for each MEM is the optimization of
$Q_\alpha$, Eq.~(\ref{eq:Q}), for a given value of $\alpha$.
In Bryan's framework, the stationarity condition $\partial Q_\alpha / \partial \v{A}= 0$ for the conventional entropy reads
\begin{equation}\label{eq:dQ}
-2 \alpha \log\frac{\v{A}}{\v{D}} = K^T \pd{\chi^2}{(K\v{A})},
\end{equation}
suggesting that a simple way to parametrize $\v{A}$ in the much smaller singular value basis is
\begin{equation}\label{eq:ansatz_A}
 \v{A} = \v{D} e^{V \v{u}},
\end{equation}
where $\v{u}$ is the new parameter vector of the same dimension as the number of kept singular values. 
With this parametrization, condition~\eqref{eq:dQ} becomes
\begin{align}
 -2 \alpha \v{u} &= \Xi U^T \pd{\chi^2}{(K\v{A})}\label{eq:condition_u}\\
     &= \Xi U^T \pd{\chi^2}{(K\v{D} e^{V \v{u}})}. \notag
\end{align}
The optimization of $Q_\alpha$ has thus been reformulated to the problem of finding the vector $\v{u}$ that solves Eq.~\eqref{eq:condition_u}, which does not explicitly depend on $\v{A}$ anymore.
This allows us to carry out the numerical solution in the (smaller) space of $\v{u}$ instead of $\v{A}$.

Ansatz~\eqref{eq:ansatz_A} ensures the positivity of $\v{A}$, so for the positive-negative approach a different parametrization has to be found in order to use the advantages of the smaller singular space.
By doing a similar derivation for $S^\pm$ as the one presented above for the conventional entropy, one realizes that due to
\begin{equation}\label{eq:derivative_entropy_pm}
\pd{S^\pm}{A_i} = -\log \frac{A_i+\sqrt{A_i^2 + 4D_i^+D_i^-}}{2D_i^+}= -\log\frac{A_i^+}{D_i^+}
\end{equation}
it is easier to express the equations in terms of $\v{A}^+$ rather than $\v{A}$.
This is possible because of relation~\eqref{eq:A-interal-opt}.
Given Eq.~\eqref{eq:derivative_entropy_pm}, a suitable parametrization is given by
\begin{eqnarray}
 \v{A}^+& = &\v{D}^+ e^{V \v{u}},\\*
 \v{A}^-& = &\v{D}^- e^{-V \v{u}},\\*
 \v{A}& = &\v{D}^+ e^{V \v{u}} - \v{D}^- e^{-V \v{u}}. \label{eq:ansatz_A_pm}
\end{eqnarray}
With this, the condition $\partial Q_\alpha / \partial \v{A}= 0$ becomes
\begin{align}
-2 \alpha \v{u} &= \Xi U^T \pd{\chi^2}{(K\v{A})}\label{eq:condition_u_pm}\\
    &= \Xi U^T \pd{\chi^2}{(K(\v{D}^+ e^{V \v{u}} - \v{D}^- e^{-V \v{u}}))}. \notag
\end{align}
Note that this expression looks nearly identical to Eq.~\eqref{eq:condition_u}.
The difference is that $\v{A}$ is parametrized in terms of $\v{u}$ by Eq.~\eqref{eq:ansatz_A} in case of non-negative functions and by Eq.~\eqref{eq:ansatz_A_pm} in the generalized case; this enters Eq.~\eqref{eq:condition_u_pm} via $\chi^2(\v{A})$ on the right-hand side. 

In principle, the actual search for a vector $\v{u}$ that fulfills Eq.~\eqref{eq:condition_u} or \eqref{eq:condition_u_pm} can be performed using any suitable numerical procedure; in practice, the Levenberg-Marquardt algorithm is usually (and also in this work) employed~\cite{marquardt_algorithm_1963}.
This program to minimize $Q_\alpha$ can be implemented very similarly
for both the standard non-negative and the here-presented positive-negative case, the only changes occur due to the generalization of Eq.~\eqref{eq:ansatz_A} to Eq.~\eqref{eq:ansatz_A_pm}.

\subsection{Poor man's matrix procedure}\label{sec:poormatrix}
As discussed before, Eq.~\eqref{eq:Gtau-from-Aw} works independently for each element of the matrices $G(\tau)$ and $A(\omega)$.
This allows us to perform the AC separately for each matrix element, using the conventional entropy, Eq.~\eqref{eq:entropy-conventional}, for diagonal elements and the modified entropy, Eq.~\eqref{eq:entropy_pm}, for the off-diagonals.

However, for physical systems, the resulting spectral function matrix has to be positive semidefinite and Hermitian, which is usually not the case when performing the AC separately for each matrix element with a flat default model.
Using a flat default model reflects the total absence of previous know\-ledge on the problem.
However, we know that a necessary condition for the positive semidefiniteness of the resulting spectral function matrix is
\begin{equation}
| A_{ll'} | \leq \sqrt{A_{ll} A_{l'l'}}.
\label{eq:condition-posdef}
\end{equation}
For example, for a problem where all diagonal elements of the spectral
function are zero at a certain frequency $\omega$,
condition~\eqref{eq:condition-posdef} implies that also all
off-diagonal elements have to be zero at this $\omega$.
Thus, once the diagonal elements have been analytically continued, this condition constitutes additional know\-ledge about the problem which might be incorporated into the MEM framework by choosing the default model for the off-diagonal elements $D_{ll'}(\omega)$ accordingly,
\begin{equation}
D_{ll'}(\omega) = \sqrt{A_{ll}(\omega)A_{l'l'}(\omega)}+\epsilon.
\label{eq:poor-man-default}
\end{equation}
Here, $\epsilon$ is a small number to prevent the default model from becoming zero, so that no division by zero occurs in the entropy term.
We will show in Secs.~\ref{sec:bethe-model} and~\ref{sec:LaTiO3} that our special choice  of the default model \eqref{eq:poor-man-default} drastically improves the results of the off-diagonal elements when they are calculated element-wise, although it does not guarantee a positive semidefinite solution.
This poor man's matrix approach is especially useful if one wants to upgrade an existing MEM code by only modifying the entropy for off-diagonal elements, as setting the default model is usually a user input.

\subsection{Full matrix formulation}\label{sec:fullmatrix}
The only way to ensure that the obtained spectral function is indeed positive semidefinite and Hermitian is by treating the matrix $A_{ab}$ as a whole.
Instead of Eq.~\eqref{eq:Q}, the functional to minimize then reads
\begin{equation}
 Q_\alpha(A) = \sum_{ab}\left[\frac12 \chi^2 (A_{ab}) - \alpha S(A_{ab})\right].
 \label{eq:matrix-Q}
\end{equation}
Here, the ordinary entropy, Eq.~\eqref{eq:entropy-conventional}, is used for the diagonal elements ($a=b$), and accordingly the modified entropy, Eq.~\eqref{eq:entropy_pm}, is used for off-diagonal elements.
One way to ensure the desired properties of $A_{ab}$ is to introduce an auxiliary matrix $B$, where $A_{ab} = \sum_c B_{ca}^* B_{cb}$.
In contrast to the parametrization of the uncoupled $A_{ab}$ described in Sec.~\ref{sec:reduction}, there is no obvious singular-space parametrization here, since $A_{ab}$ couples different elements of $B$.
However, as the elements $B_{ab}$ can be positive and negative for both diagonal and off-diagonal elements, in the spirit of Sec.~\ref{sec:reduction} we choose
\begin{equation}
\v{B}_{ab} = \v{D}_{ab} \left(e^{V \v{u}_{ab}} -  e^{-V \v{u}_{ab}}\right).
    \label{eq:param-B}
\end{equation}
Using the resulting parametrization of $\v{A}_{ab}$ in terms of the
singular-space vectors $\v{u}$, the stationarity condition for
$Q_\alpha$ from Eq.~\eqref{eq:matrix-Q} leads to equations
which consequently have to be solved for $\v{u}$ (for a more detailed discussion, see Appendix~\ref{sec:statcond}).
The fact that the expression for $A_{ab}$ now couples the singular-space parameters $\v{u}$ of different matrix elements means that all matrix elements have to be treated at the same time.
Consequently, the configuration space grows quadratically with the matrix size $d$.
Concerning the computational cost, the fundamental difference between the poor man's and the full matrix approach is that in the former, one needs to $d^2$ times find a solution in a configuration space of size $N_\Xi$, while in the latter, one searches a solution once in a configuration space of size $N_\Xi\cdot d^2$.
As typically solver algorithms take disproportionally longer for larger search spaces, this usually leads to a substantial increase of computational time.
Nevertheless, the increased computational effort is justified, as it gives the possibility to ensure the desired properties of $A$, leading to a large improvement of quality. 
A flat default model is chosen for all matrix elements when the full matrix method is used in this paper.

\subsection{Analytic continuation of the self-energy}\label{sec:ac_sigma}
One of the central quantities of many-body theory is the self-energy $\Sigma$.
While some of its properties can be understood from $\Sigma(i\omega_n)$, the analytically continued $\Sigma(\omega+i0^+)$ allows a more straightforward interpretation and the calculation of further physical properties.

We will focus our discussion of the AC of the self-energy on DMFT~\cite{DMFT2}, where the self-energy is approximated to be $k$~independent and connects the impurity to the lattice problem.
For a given (in general, matrix-valued) $\Sigma$, the local
(matrix-valued) lattice Green's function is
\begin{equation}
    G_{loc}(z) = \sum_k [z - \mu - H_k - \Sigma(z)]^{-1}.
    \label{eq:gloc-dmft}
\end{equation}
The matrix $H_k$ is the $k$-dependent Hamiltonian of the lattice and the inversion has to be understood as matrix inversion.
The so-called impurity Weiss field $\mathcal{G}_0(z)$ is obtained from Dyson's equation
\begin{equation}
    \mathcal{G}_{0}^{-1}(z) = G^{-1}_{loc}(z) + \Sigma(z).
\end{equation}
This $\mathcal{G}_0$ is the input for the impurity solver to calculate the self-energy and the interacting impurity Green's function $G_{imp}$; when inserting $\Sigma$ back into Eq.~\eqref{eq:gloc-dmft}, the self-consistency loop can be closed.
The DMFT cycle is iterated until convergence is reached, i.e.,\ until $G_{loc} = G_{imp}$.
The self-energy as a function of real frequency is needed within the framework of DMFT to calculate lattice quantities, e.g.,\ $G_{loc}(\omega+i0^+)$ as defined in Eq.~\eqref{eq:gloc-dmft}, $k$-resolved spectral functions~\cite{liebsch_photoemission_2000,biermann_observation_2004}, Fermi surfaces~\cite{liebsch_photoemission_2000} or optical properties of strongly correlated materials~\cite{optic_oudovenko}.

In contrast to Green's functions, there is no relation equivalent to Eq.~\eqref{eq:spectral-representation} for self-energies and, hence, one needs to find an appropriate method to perform the AC.
There are several ways to do so.
One could analytically continue both the DMFT Weiss field $\mathcal{G}_0(i\omega_n)$ and the interacting impurity Green's function $G_{imp}(i\omega_n)$ and calculate $\Sigma(\omega+i0^+)$ via the Dyson equation on the real-frequency axis~\cite{wang_antiferromagnetism_2009}.
However, there are two independent analytic continuations involved, and hence, the resulting real-frequency self-energy tends to oscillate heavily and does usually give poor results (see, e.g.,\ Ref.~\cite{wang_antiferromagnetism_2009}).
Another approach is to solve for $\Sigma(\omega+i0^+)$ in the expression for the (analytically continued) $G_{loc}(\omega+i0^+)$~\cite{jarrell_optical_1995, anisimov_full_2005, tomczak_effective_2007}.

The most commonly used approach in literature is to continue an auxiliary quantity $G_{aux}$.
Overall, this requires the following five steps: (i) construction of $G_{aux}(i\omega_n)$ from the self-energy $\Sigma(i\omega_n)$,
(ii) inverse Fourier transform~\footnote{%
In order to avoid spurious oscillations in the inversely Fourier-transformed $G(\tau)$, we fit the high Matsubara frequencies of $\Sigma(i\omega_n)$ with its high-frequency expansion in $\omega_n$ (``tail fit'') and subtract the resulting tail from $G_{aux}$ before performing the inverse Fourier transform. Then we add the analytic inverse Fourier transform of the tail expansion on the $\tau$~axis.%
} to $G_{aux}(\tau)$, 
(iii) AC of $G_{aux}(\tau)$ to $A_{aux}(\omega)$,
(iv) construction of $G_{aux}(\omega+i0^+)$ from $A_{aux}(\omega)$ using Eq.~\eqref{eq:spectral-representation}, and finally
(v) obtaining $\Sigma(\omega+i0^+)$ from $G_{aux}(\omega+i0^+)$.

In the following, we give two possible constructions of $G_{aux}(z)$.
First, one can use
\begin{equation}
G_{aux}(z) = \Sigma(z) - \Sigma(i\infty)
\label{eq:G-aux-cor}
\end{equation}
where $\Sigma(i\infty)$ is the constant term of the high-frequency expansion of $\Sigma(i\omega_n)$~\cite{wang_antiferromagnetism_2009}. 
We note here that the resulting quantity $G_{aux}$ is, technically speaking,
not a Green's function, since its off-diagonal elements do not have the correct analytic high-frequency behavior (they should fall off like $\sim 1/(\omega_n)^2$, but in Eq.~\eqref{eq:G-aux-cor} they fall off like $\sim 1/\omega_n$).
Second, there is the inversion method, e.g.,\ used in Ref.~\cite{jernej_invsigma},
\begin{equation}
G_{aux}(z) = [z + C - \Sigma(z)]^{-1}.
\end{equation}
The constant $C$ is usually set to $C= \Sigma(i\infty) + \mu$ with the chemical potential $\mu$.
In this work, we choose to use the inversion method.

\subsection{Implementation details}
\label{sec:implementation}
We implement a variation of Bryan's MEM algorithm~\cite{bryan_1990} allowing arbitrary expressions for the entropy with the ability to treat the problem in the full matrix formulation.
For the minimum search of $Q_\alpha$ we use the Levenberg-Marquardt minimization algorithm~\cite{marquardt_algorithm_1963}.
The expressions for the step length and the convergence criterion are chosen as in Ref.~\cite{bryan_1990}.
The spectral function is parametrized in singular space as laid out in Secs.~\ref{sec:reduction} and~\ref{sec:fullmatrix} and the value of the hyperparameter $\alpha$ is chosen using the piecewise linear fit of $\log\chi^2(\log \alpha)$ (see Sec.~\ref{sec:hyper}).
In general, the frequency mesh on which $A(\omega)$ is discretized can be freely chosen.
In this work, we use a hyperbolic grid, which asymptotically becomes a linear grid for high frequencies but is denser around $\omega=0$.
This allows the use of a smaller overall number of $\omega$~points, which speeds up the calculation.
However, when calculating the full Green's function from the spectral function according to Eq.~\eqref{eq:spectral-representation}, a small broadening ($i0^+$) has to be used.
As the hyperbolic grid with a small number of points is not dense enough (we use $N_\omega = 200$ points), this small broadening leads to artefacts, which can be avoided by first interpolating the spectral function on a much finer grid (we use a linear mesh with $10\,000$ points) and then using $2\Delta\omega$ of the new grid as broadening.

For metallic systems, it is known that the MEM spectra tend to exhibit spurious cusps around the Fermi level~\cite{silver_maximum-entropy_1990}.
This can be prevented by using the preblur formalism~\cite{skilling_preblur}, where the so-called ``hidden spectral function'' is blurred via a convolution with a Gaussian function. 
Within this algorithm, the hidden spectral function is used to calculate the entropy $S$, but the misfit $\chi^2$ is evaluated from the blurred spectral function.
Also, this blurred spectral function is what is taken in the end as the solution of the problem.

The width $b$ of the Gaussian is another hyperparameter that can be chosen similarly to $\alpha$, e.g.,\ by searching the maximum of the probability $p(\alpha, b)$ or by locating the characteristic kink in $\log\chi^2(\log \alpha, \log b)$.
In accordance to the route employed for determining $\alpha$, first we determine one value of $\alpha$ for every value of $b$ using the fit method described at the end of Sec.~\ref{sec:hyper}.
Then, we take the curves of $\log \chi^2$ at that value of $\alpha$ for the different values of $b$ and fit once more to determine $b$.

\section{Two-band model}
\label{sec:bethe-model}
As a benchmark system for the presented approach, we investigate an artificial particle-hole symmetric two-band model with semicircular density of states. We set the half-band width to $D_1 = 2$ for the first band and to $D_2 = 1$ for the second band.
We choose the interaction term in a simple Hubbard-type form $H_{int} = \sum_i U_i n_{i\uparrow} n_{i\downarrow}$ with $U_i/D_i = 3.25$.
The chemical potential and the on-site energies are chosen such that both bands are half-filled.
For the chosen interaction, the system is a Mott insulator with a spectrum consisting of two distinct Hubbard bands separated by an energy $U_i$.
We treat the problem with DMFT to obtain an interacting impurity Green's function $G_{imp}(\tau)$ and a self-energy $\Sigma(i\omega_n)$ at an inverse temperature of $\beta D_2= 40$.
The simplicity of the problem allows the use of iterated perturbation theory (IPT)~\cite{yosida_perturbation_1970, yamada_perturbation_1975, yosida_perturbation_1975, rozenberg_mott-hubbard_1994, DMFT2} as impurity solver.
The AC of the IPT results are performed using Pad\'e approximants~\cite{ferris-prabhu_numerical_1973,beach_reliable_2000}.
Because of the noiseless nature of the IPT data, the Pad\'e approximants give reliable results for this specific problem~\cite{DMFT2}.
Additionally we solve our two-band model using the continuous-time hybridization-expansion quantum Monte Carlo solver (\textsc{triqs/cthyb}) ~\cite{TRIQS/CTHYB,werner2}, which is based on the \textsc{triqs} package~\cite{TRIQS}, and perform the AC with the MEM.
We perform \num{8e6} CTHYB measurements. 
Of course, more measurements would undoubtedly be beneficial for the AC. 
Nevertheless, we limit ourselves here to emulate more complicated situations where higher-quality data can only be obtained with a substantial increase in computational effort.
Although it would be possible to evaluate the covariance matrix of the
Monte Carlo to take into account the correlations of the noise of
$G(\tau)$ at different values of $\tau$, for simplicity, we estimate
the Monte Carlo noise by manual inspection of the imaginary-time data
($5\times 10^{-4}$ in our case), and we assume a diagonal covariance matrix with a constant noise for these (and the following) tests.
As we determine $\alpha$ by detecting the characteristic kink in $\log\chi^2(\log \alpha)$, the procedure is less sensitive to the given error than, e.g., the classic MEM (see Sec.~\ref{sec:hyper}).

In the following, we will compare the curves obtained by IPT and Pad\'e with those from CTHYB and MEM as two approaches to tackle this problem.
The former suffers from a systematic error as it is a perturbative technique, but yields results without statistical error.
The latter, on the other hand, is exact in theory, but will always give noisy Green's functions and, thus, uncertainties after AC.
In context of multiorbital DMFT calculations away from half-filling, in many cases quantum Monte Carlo impurity solvers are the only option, making it necessary to analytically continue noisy data.

Some tests benchmarking our implementation of the MEM algorithm and an investigation of the effect of random noise on the data can be found in Appendix~\ref{sec:appl-bench}.

In order to model a system with off-diagonal Green's functions and
self-energies, we perform a basis transformation for the
$G_{imp}(\tau)$ and $\Sigma(i\omega_n)$, which come out as diagonal
matrices from the impurity solvers.
In this work, we simply use a rotation matrix with an angle $\phi = \SI{0.4}{\radian}$, which is representative for the results obtained for other angles.

\begin{figure}[t]
    \centering
    \includegraphics[width=\linewidth]{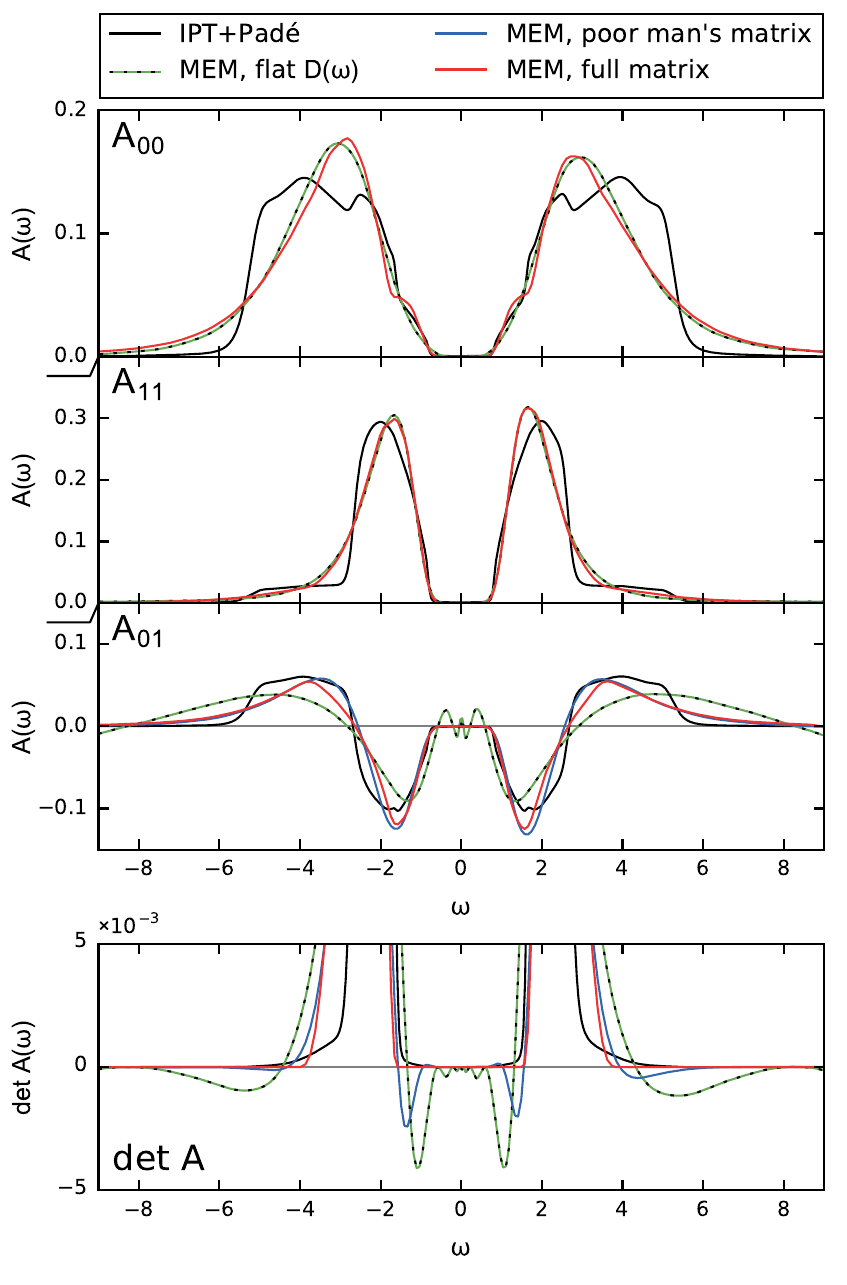}
    \caption{
    Top:
    Spectral function of the rotated model in the insulating regime.
    Each subplot represents one matrix element, the two off-diagonal elements $A_{01}$ and $A_{10}$ being the same.
    The result from IPT and Pad\'e (black) is shown along with the MEM results.
    For the latter, we compare the continuation treating the matrix elements independently with a flat default model (dashed green) with the poor man's matrix method, i.e.\ using a default model incorporating the information from the diagonal elements (blue, only for the off-diagonal elements).
    Furthermore, the result of the matrix formulation (red) that ensures a positive semidefinite, Hermitian spectral function is shown.
    Bottom:
    The determinant of the matrix-valued spectral function $A$ as a function of frequency.
    Wherever $\det A(\omega)$ is negative, the matrix is not positive semidefinite.
    }
    \label{fig:model-gloc-comp}
\end{figure}

Figure~\ref{fig:model-gloc-comp} shows the resulting spectral function for the AC of the rotated $G_{imp}(\tau)$.
Using a flat default model, the off-diagonal elements of the spectral function feature strong oscillations (dashed green line).
This can be explained by the relaxation of the positivity constraint: 
In general, the AC tends to overfit around $\omega = 0$ and to underfit for large $\omega$, since the kernel is largest for small $\omega$. 
For metallic spectral functions, these artefacts can be cured as explained in Sec.~\ref{sec:implementation}.
For insulating spectral functions, the oscillations around $A=0$ are suppressed in the diagonal components, because fluctuations to negative values are not possible due to their positivity. 
In the off-diagonal elements, however, these fluctuations appear. 
Additionally, for high frequencies, the solution for the off-diagonal elements with flat default model does not tend to zero as in the IPT and Pad\'e solution, but overshoots and goes to negative values.
The violation of the particle-hole symmetry is due to the stochastic nature of the Monte Carlo data (it is thus a measure for the quality of the QMC result) and not directly a fault of the MEM (see Appendix~\ref{sec:appl-bench}).
We refrained from symmetrizing the resulting $G(\tau)$, because in most models one does not have this possibility.

\begin{figure}
    \centering
    \includegraphics[width=\linewidth]{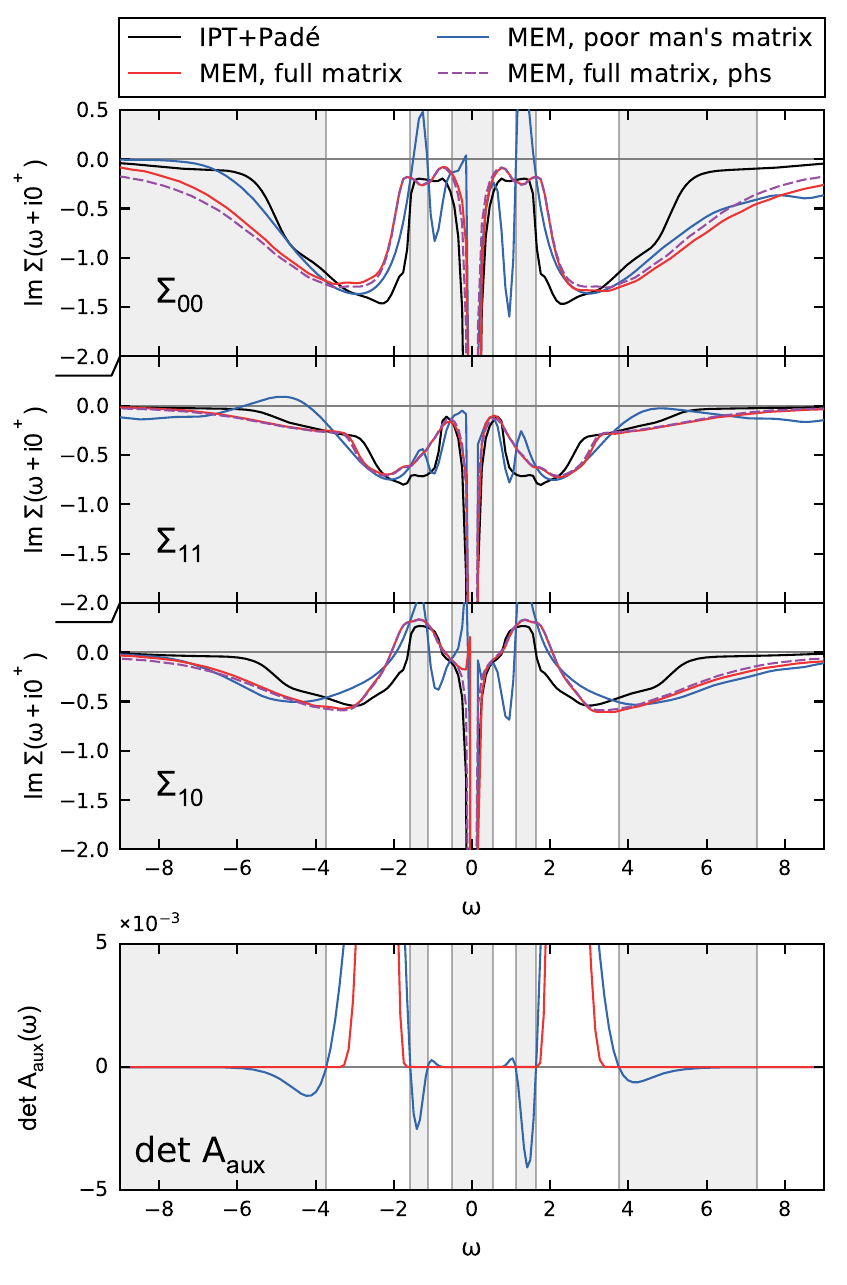}
    \caption{
    Top:
    Imaginary part of the self-energy $\Sigma(\omega+i0^+)$ for the rotated model obtained with the inversion method.
    Each subplot re\-pre\-sents one matrix element, the two off-diagonal elements $\Sigma_{01}$ and $\Sigma_{10}$ being the same.
    The result from IPT and Pad\'e (black) is compared to the curves obtained with CTHYB and the MEM.
    The poor man's matrix method (blue) is presented alongside the full matrix method (red).
    For the latter, also the result where the auxiliary spectral function, $A_{aux}$, has been particle-hole symmetrized (phs) is shown (dashed purple line).
    Bottom:
    The determinant of the matrix-valued auxiliary spectral function $A_{aux}$ as a function of frequency.
    Wherever $\det A_{aux}(\omega)$ is negative, the matrix is not positive semidefinite.
    This happens only for the poor man's matrix method.
    The regions where the corresponding $A_{aux}$ is not positive semidefinite are marked by the gray areas.
    }
    \label{fig:model-sigma-comp}
\end{figure}

Building the information from the diagonal elements into the default model of the off-diagonal elements (poor man's method), as outlined in Sec.~\ref{sec:poormatrix}, improves these issues (blue line in Fig.~\ref{fig:model-gloc-comp}).
It suppresses oscillations of the off-diagonal elements where the diagonal elements are small, stabilizes a smooth solution, and improves the high-frequency behavior.
Nevertheless, care has to be taken as this does not mean that the solution is positive semidefinite, and indeed, at some frequencies that general property of the spectral function is violated (see the plot of $\det A$ at the bottom of Fig.~\ref{fig:model-gloc-comp}).
Therefore, we apply the full matrix formulation (Sec.~\ref{sec:fullmatrix}) to the problem (red lines in Fig.~\ref{fig:model-gloc-comp}).
This solves the issues one faces when performing the AC separately for the individual matrix elements.
The spurious oscillations of the off-diagonal elements of $A(\omega)$ are efficiently suppressed even for a flat default model and the spectral function matrix is positive semidefinite everywhere.
\begin{figure}
    \centering
    \includegraphics[width=\linewidth]{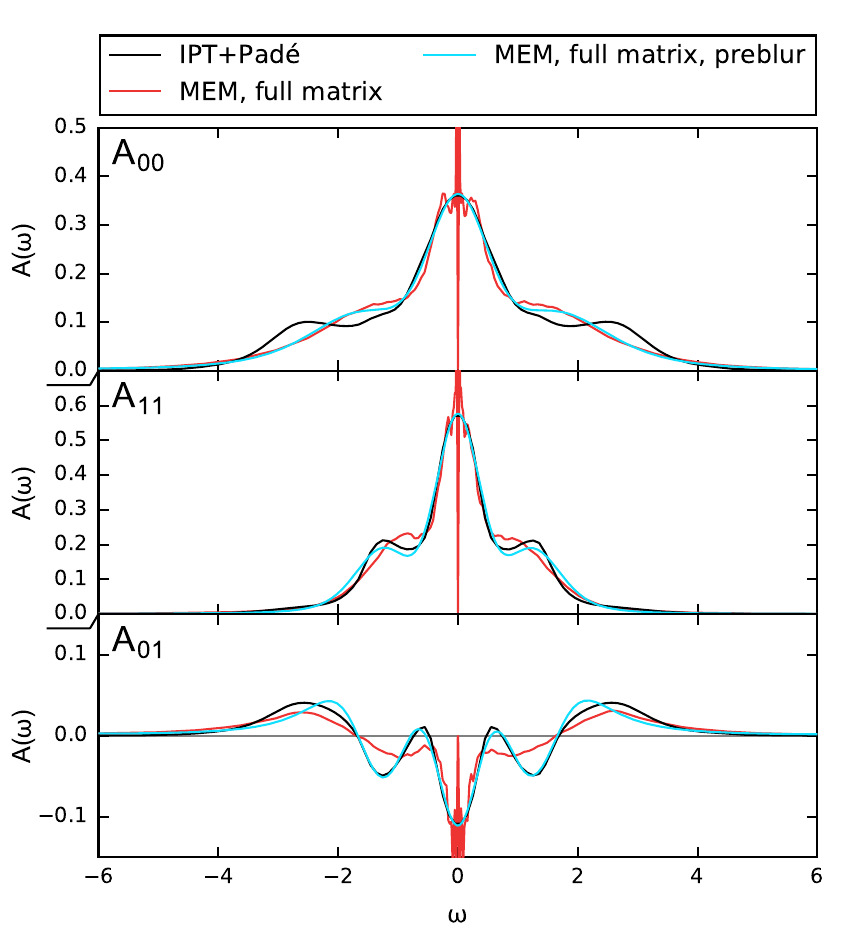}
    \caption{Spectral function of the rotated model in the metallic regime.
    Each subplot represents one matrix element, the two off-diagonal elements $A_{01}$ and $A_{10}$ being the same.
    The result from IPT and Pad\'e (black) is shown along with the full matrix MEM result.
    For the latter, we show the result with (cyan) and without (red) using the preblur method.
    }
    \label{fig:preblur-gloc-plot}
\end{figure}

As a next step, we benchmark the AC of $\Sigma$, using the inversion method to construct an auxiliary Green's function (see~Sec.~\ref{sec:ac_sigma}) for the off-diagonal model; the obtained $\Sigma(\omega+i0^+)$ is shown in Fig.~\ref{fig:model-sigma-comp}.
The separate AC of the individual matrix elements using a flat default model leads to a heavily oscillating self-energy, which is why it is not shown here.
But even performing the poor man's matrix method (blue line in Fig.~\ref{fig:model-sigma-comp}) leads to unphysical results.
Especially in the regions where the auxiliary spectral function $A_{aux}$ is not positive semidefinite (shaded in gray in Fig.~\ref{fig:model-sigma-comp}), these problems are evident:
there are heavy oscillations when the curve overshoots whenever the derivative changes quickly, and for some frequencies even the diagonal elements of the imaginary part of the self-energy become positive.
This shows that the poor man's method is not adequate for determining a matrix-valued $\Sigma(\omega+i0^+)$.
The full matrix formulation (red line in Fig.~\ref{fig:model-sigma-comp}), however, yields physical solutions just as IPT and Pad\'e; these two solutions are consistent with each other within the limits of the method.
Again, the slight deviation from the particle-hole-symmetrized result (dashed purple line) is due to a stochastic violation of that symmetry in the $G(\tau)$ data.
This also manifests itself as a spurious peak close to $\omega=0$ in the off-diagonal element of $\Sigma(\omega+i0^+)$, which can be traced back to a slight mismatch of the position of the poles of $\operatorname{Im}\Sigma(\omega+i0^+)$ (that should be at $\omega=0$) between different matrix elements.
In general models, the particle-hole symmetry is not present and cannot therefore be exploited to improve the result.
The real part of $\Sigma(\omega+i0^+)$, which is related to the imaginary part by the Kramers-Kronig relation, also gives plausible results and the two different methods agree very well (not shown).
Once $\Sigma(\omega+i0^+)$ has been obtained, other lattice quantities are accessible.
However, we do not further discuss this here, but refer the reader to the example in the next section (Sec.~\ref{sec:LaTiO3}), where we also calculate the local lattice Green's function $G_{loc}$ from $\Sigma(\omega+i0^+)$.

So far, only an insulating solution has been investigated.
However, we also put the method to a test in the metallic regime of the model ($U_i/D_i = \num{1.5}$), shown in Fig.~\ref{fig:preblur-gloc-plot}.
As discussed in Sec.~\ref{sec:implementation}, it is necessary to use preblurring to avoid cusps around $\omega=0$.
This is also observed in the generalization of the method to off-diagonal elements.
Not only is the preblurred spectral function smoother around $\omega=0$, but also more details at higher frequencies can be resolved, which can be best seen in the off-diagonal element.
In general, in that regime, the results from CTHYB and MEM (employing the preblur technique) and IPT and Pad\'e agree very nicely, similar to what is found for the insulating case.

\section{\label{sec:LaTiO3} Application: $\mathbf{LaTiO_3}$ }

\begin{figure}
\centering
\includegraphics[width=0.95\linewidth]{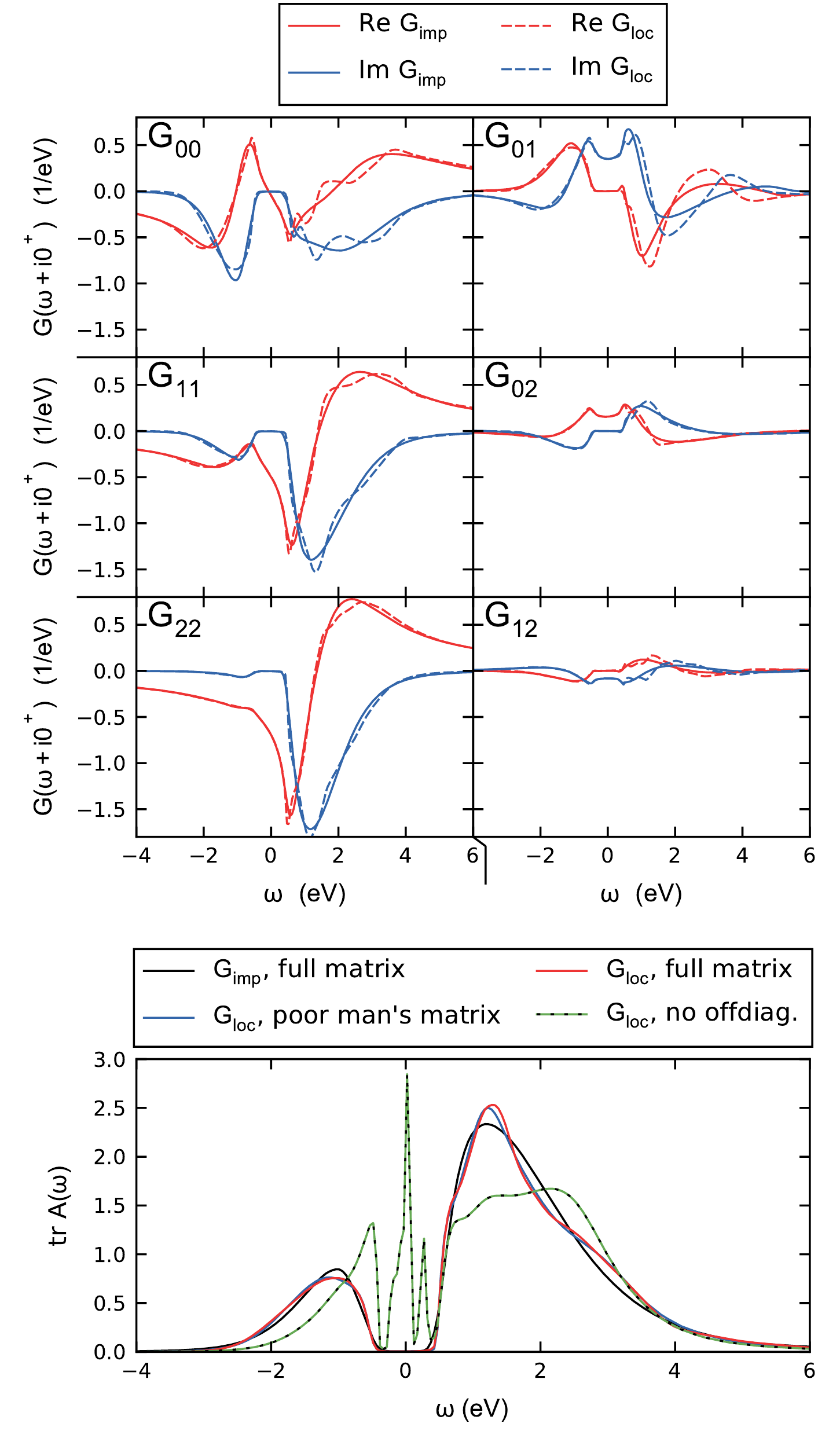}
\caption{\label{fig:LaTiO3} Top: Comparison of real (red) and imaginary parts (blue) of the impurity Green's function $G_{imp}(\omega+i0^+)$ (solid lines) and the local lattice Green's function $G_{loc}(\omega+i0^+)$ (dashed lines).
The former is obtained by a direct AC, whereas the latter is calculated via Eq.~\eqref{eq:gloc-dmft} after the AC of the self-energy.
In both cases, the matrix formulation of the MEM code was used.
The subplots represent different matrix elements of the Green's function.
\\
Bottom:
Total spectral function (i.e., trace over the orbital and spin degrees of freedom) of the Ti-$t_{2g}$ bands from the Green's functions shown above ($G_{imp}$, black, and $G_{loc}$, red).
For $G_{loc}$, performing the AC of $\Sigma$ using the poor man's matrix method is shown as well (blue).
Additionally, we show the spectral function of a local Green's function where we have set the off-diagonal elements of the self-energy $\Sigma(i\omega_n)$ to zero before individually continuing its diagonal elements and evaluating $G_{loc}$ from the obtained $\Sigma(\omega+i0^+)$ (dashed green).
\\\\\\\\~
}
\end{figure}

Finally, we apply the matrix formulations presented above in Secs.~\ref{sec:poormatrix} and~\ref{sec:fullmatrix} to \ce{LaTiO3}, for which we perform a one-shot DFT+DMFT calculation.
The transition metal oxide \ce{LaTiO3} has a perovskite crystal structure with tilted oxygen octahedra and distorted lanthanum cages.
Because of these structural distortions, the material features an off-diagonal hybridization, and thus also an off-diagonal impurity Green's function $G_{imp}(\tau)$. \ce{LaTiO3} was already extensively analyzed in literature~\cite{perov_Pavarini,craco,LaTiO3_Pavarini}, where also the nature of the Mott insulating state was traced back to the tilting and rotation of the oxygen octahedra and the accompanying lifting of the $t_{2g}$ degeneracy.

Here, we do not further elaborate on the physics, but rather use \ce{LaTiO3} as a benchmark material to prove the following points: 
First, we emphasize that the analytic continuation of off-diagonal elements is a problem often encountered in real-materials calculations.
Second, the calculations presented here show that the full matrix formalism is feasible for $3 \times 3$ matrices.
Third, we show that the continuation of the self-energy leads to a local Green's function $G_{loc}(\omega+i0^+)$ which is comparable to the continuation of $G_{imp}$.

Our calculations were carried out with \textsc{wien2k}~\cite{Wien2k1} and the \textsc{triqs/dfttools} package~\cite{TRIQS/DFTTOOLS,LaFe01,TRIQS/DFTTools2,TRIQS}.
For the DFT part, we use the crystal structure from Ref.~\cite{LaTiO3_struct}, \num{40000} $k$~points in the full Brillouin zone and employ the standard Perdew-Burke-Ernzerhof (PBE)~\cite{PBE} generalized gradient approximation (GGA) for the exchange-correlation functional.
From the DFT Bloch states we construct projective Wannier functions for the $t_{2g}$ subspace of the \mbox{Ti-$3d$} states in an energy window from \num{-1.0} to \SI{1.2}{\electronvolt} around the Fermi level.
In DMFT, we use the Kanamori Hamiltonian with a Coloumb interaction $U=\SI{4.5}{\electronvolt}$ and a Hund's coupling $J=\SI{0.65}{\electronvolt}$ similar to the values used in Ref.~\cite{LaTiO3_Dang, LaTiO3_Pavarini}.
We solve the impurity model on the imaginary axis with the \textsc{triqs/cthyb} solver~\cite{TRIQS/CTHYB} at an inverse temperature $\beta= \SI{40}{\electronvolt^{-1}}$ and use a total number of \num{3.2e7} measurements.
We choose the solver basis such that the density matrix is diagonal. In the case of \ce{LaTiO3}, this basis has the advantage that all matrix elements of $G_{imp}(\tau)$ are real if the phases are chosen accordingly.

Having obtained $G_{imp}(\tau)$ from the DFT+DMFT calculation, the AC is again performed in two ways:
First, with the full matrix formalism for the full Green's function matrix (Sec.~\ref{sec:fullmatrix}) and second, by a separate continuation of the individual elements with the poor man's matrix method introduced in Sec.~\ref{sec:poormatrix}. Furthermore, we analytically continue $\Sigma(i\omega_n)$ by means of the inversion method (see Sec.~\ref{sec:ac_sigma}).
We calculate the local Green's function $G_{loc}(\omega+i0^+)$ with Eq.~\eqref{eq:gloc-dmft} and compare it to the direct continuation of the impurity Green's function $G_{imp}(\omega+i0^+)$ in the top graph of Fig.~\ref{fig:LaTiO3}.

Within DMFT, the self-consistency condition requires $G_{loc} = G_{imp}$, which is well fulfilled on the Matsubara axis.
Nevertheless, the agreement on the real axis shown in Fig.~\ref{fig:LaTiO3} is remarkably good for both the diagonal and the off-diagonal elements, especially when considering the different magnitudes of the individual matrix elements and the fact that the continuation is performed for different Green's functions, i.e.,\ $G_{imp}(i\omega_n)$ and $G_{aux}(i\omega_n)$. This underlines the capabilities of the presented full matrix method. 

Here, we only show the Green's functions obtained with the full matrix formalism; however, it should be emphasized that for \ce{LaTiO3} also the poor man's method gives very similar results (see the corresponding spectral function in the bottom graph of Fig.~\ref{fig:LaTiO3}). Therefore, here the element-wise continuation with the poor man's method constitutes an efficient alternative to the full matrix method.

Figure~\ref{fig:LaTiO3} does not only prove the concept of the AC for the full Green's function, but also shows that the AC of the self-energy via the construction of an auxiliary Green's function is a feasible approach.
In contrast, calculating the spectral function from $G_{loc}(\omega+i0^+)$, where we set the off-diagonal elements of the self-energy $\Sigma(i\omega_n)$ to zero (and thus analytically continue only the diagonal elements), does lead to a completely wrong, even metallic, spectral function (see dashed green line in bottom plot of Fig.~\ref{fig:LaTiO3}).
This clearly shows that the off-diagonal elements must not be neglected at this point of the calculation.

In terms of the gap as well as the overall shape and size of the Hubbard bands, the presented spectra for the Ti-$t_{2g}$ subspace are in good agreement with calculations available in the literature~\cite{LaTiO3_Pavarini, perov_Pavarini, LaTiO3_Dang}.

\section{\label{sec:Conc}Conclusion}

In this work, we show how a consistent framework for the analytic
continuation of matrix-valued Green's functions can be constructed on
a probabilistic footing. 
In order to enable a treatment of the off-diagonal elements, we use an
entropy that allows to relax the non-negativity constraint one has to
obey in the usual maximum entropy method.
With this generalization, diagonal and off-diagonal elements can, in
principle, be treated on a similar footing.  

The practical use of this method is studied on two examples, an
artificial two-band model and a realistic DFT+DMFT calculation for the
insulating compound \ce{LaTiO3}.
First, we propose the poor man's matrix method, where the matrix
elements are treated separately. With this scheme, we find
satisfactory results for some cases (e.g.,\ for \ce{LaTiO3}), but also
see completely unphysical results in our calculations for the two-band
model, since positive semidefiniteness and Hermiticity of the
spectral functions cannot be guaranteed.

Only the AC in full matrix formulation cures these problems and
produces spectral functions with the correct mathematical
properties, such as positive semidefiniteness and Hermiticity.
Although being computationally more expensive, it should be employed
whenever feasible. 

Moreover, these methods for AC introduced here give access to the
matrix-valued self-energy on the real-frequency axis, which is
indispensable for the study of lattice quantities. 

\section*{\label{sec:Ackn}Acknowledgments}
The authors want to thank E.~Schachinger for fruitful discussions and acknowledge financial support from the Austrian Science Fund, Projects No.\ Y746, No.\ P26220, and No.\ F04103.
The computational results presented have been achieved in part using the Vienna Scientific Cluster (VSC).
The \textsc{python} libraries numpy~\cite{numpy} and matplotlib~\cite{matplotlib} have been used.

\appendix

\section{Probabilities}\label{sec:prob}

\subsection{Conventional entropy}
The framework presented here was developed in the pioneering works by Skilling~\cite{skilling_1989}, Gull~\cite{gull_1989} and Bryan~\cite{bryan_1990}, but we will rephrase it here for completeness.
In Ref.~\cite{skilling_1989}, Skilling used the picture of the monkeys presented in Sec.~\ref{sec:posneg} to relate the entropy to a prior probability distribution
\begin{eqnarray}
 P(\v{A}|\v{D},\alpha)& = &\frac{1}{Z_S} \; e^{\alpha S(\v{A},\v{D})}, \label{eq:entropic-prior}\\*
 Z_S& = &\int \frac{d^N A}{\prod_i \sqrt{A_i}} \; \; e^{\alpha S(\v{A},\v{D})}. \nonumber
\end{eqnarray}
Note that the measure in Eq.~\eqref{eq:entropic-prior} is not flat, but $\prod_i A^{-1/2}_i$.
The metric of the spectral function space is therefore $g_{ij} = \delta_{ij}/A_i$, which is minus the second derivative of the entropy $g_{ij} = -\partial^2 S/ \partial A_i \partial A_j$~\cite{levine_1986, skilling_1989, rodriguez_1989}.
The prior distribution is maximized by $A_i = D_i$. When expanding the entropy to second order around the maximum, the variance is therefore $\alpha D_i$.
Thus, the default model determines both the maximum as well as the width of the distribution, as expected from the assumed Poisson process.

Using the likelihood $P(\v{G}|\v{A}) =  e^{-\chi^2/2}/Z_{\chi^2}$ with $Z_{\chi^2}= \int d^N G\; e^{-\chi^2/2}$, the minimization of $Q_\alpha$ [see Eq.~\eqref{eq:Q}] can be understood as a maximization of the probability $P(\v{G},\v{A}|\alpha,\v{D}) = e^{-Q_\alpha}/(Z_S Z_{\chi^2})$.
This can be used to determine $\alpha$ on a probabilistic footing since marginalizing over $\v{A}$ gives~\cite{gull_1989}
\begin{equation} \label{eq:p_alpha}
 P(\alpha | \v{G},\v{D}) \propto P(\alpha) \int \frac{d^N A}{\prod_i \sqrt{A_i}} e^{-\frac{1}{2}\chi^2 + \alpha S}.
\end{equation}
The prior $P(\alpha)$ is usually chosen to be Jeffrey's prior $1/\alpha$.
The most common way to evaluate the integral~\eqref{eq:p_alpha} is to expand the exponent up to second order.
The final expression for the probability is~\cite{gull_1989}
\begin{equation}\label{eq:p_alpha2}
 P(\alpha|\v{G},\v{D}) \propto P(\alpha) \alpha ^\frac{N_\tau}{2} e^{-Q_\alpha(\v{A}^*_\alpha)}\left(\det(\Lambda+ \alpha)\right)^{-\frac12},
\end{equation}
where $\v{A}^*_\alpha$ minimizes $Q_\alpha$ and $\Lambda$ is a matrix with the elements $\Lambda_{ij} = \frac12  \sqrt{A_i A_j}\,\partial^2\chi^2/\partial A_i \partial A_j$.
The classical MEM by Gull uses the $\alpha$ that maximizes $ P(\alpha | \v{G},\v{D}) $ within this approximation~\cite{gull_1989}, whereas Bryan suggested to calculate the weighted average $\bar{\v{A}} = \int d\alpha P(\alpha | \v{G},\v{D}) \v{A}_\alpha$~\cite{bryan_1990}.
In most cases, $P(\alpha | \v{G},\v{D})$ is sharply peaked, and thus, the classical and the Bryan MEM give very similar results.\\

\subsection{Positive-negative entropy}

Since $\v{A}^+$ and $\v{A}^-$ are assumed to be independent, the entropy is $S(\v{A}^+,\v{A}^-) = S(\v{A}^+)+ S(\v{A}^-)$ and one has to marginalize over both $\v{A}^+$ and $\v{A}^-$ in order to find the probability of $\alpha$:
\begin{eqnarray} \label{eq:p_alpha_pm}
 P(\alpha | \v{G},\v{D}) & \propto & P(\alpha) \int \frac{d^N A^+}{\prod_i \sqrt{A^+_i}} \int \frac{d^N A^-}{\prod_i \sqrt{A^-_i}}\\*
 & &\times \;e^{-\frac{1}{2}\chi^2(\v{A}=\v{A}^+ - \v{A}^-) + \alpha S(\v{A}^+,\v{A}^-)}. \nonumber
\end{eqnarray}
As in the derivation of $S^\pm$, one can use the fact that $\chi^2$ depends only on the difference $\v{A}^+-\v{A}^-$, so that the remaining degree of freedom can be integrated out. Transforming to $\v{A}= \v{A}^+-\v{A}^-$ and some auxiliary $\v{A}' = \v{A}^++\v{A}^-$, the integral over $\v{A}'$ is easily evaluated using a second-order expansion of the exponent like in the classical MEM, yielding
\begin{equation} \label{eq:p_alpha_pm2}
 P(\alpha | \v{G},\v{D}) \propto P(\alpha) \int \frac{d^N A}{\prod_i \sqrt[4]{A_i^2 + 4D_i^+D_i^-}} e^{-\frac{1}{2}\chi^2 + \alpha S^\pm}.
\end{equation}
The same form is also obtained in Ref.~\cite{hobson_1998}, using a different approach to prove it, with the Skellam distribution of the two monkey hordes as a starting point.
Note that either way Eq.~\eqref{eq:p_alpha_pm2} is an approximation.
This expression looks similar to the case of positive spectral functions~\eqref{eq:p_alpha}, but with a different entropy given by Eq.~\eqref{eq:entropy_pm} instead of the ordinary entropy~\eqref{eq:entropy-conventional} and with the measure $\prod_i(A_i^2 + 4D_i^+D_i^-)^{-1/4}$ instead of $\prod_i A_i^{-1/2}$.
Interestingly, the metric is given by the second derivative of the entropy $g_{ij} = -\partial^2 S/ \partial A_i \partial A_j$, just as in the ordinary case of non-negative spectral functions.
Expanding the exponent of integral~\eqref{eq:p_alpha_pm2} again to second order, the probability has exactly the same form as in the strictly positive case~\eqref{eq:p_alpha2}, but with a different $Q_\alpha = \frac12 \chi^2 - \alpha S^\pm$ due to the different entropy and with a different matrix  $\Lambda_{ij} =  \frac12 \sqrt[4]{A_i^2+ 4 D_i^+D_i^-}\partial^2\chi^2/\partial A_i \partial A_j \sqrt[4]{A_j^2+ 4 D_j^+D_j^-}$ due to the different metric. The prior distribution $e^{\alpha S^\pm(\v{A})}$ is maximized by $A_i = D_i^+ - D^-_i$. Expanding the entropy up to second order around this maximum gives a variance of $\alpha(D_i^+ + D_i^-)$. Thus, in the usual case $D^+_i = D^-_i = D_i$, the default model only influences the solution via the width of the distribution, not via the position of the maximum, since this is always at $A_i = 0$.

\section{Stationarity condition in the full matrix formalism}
\label{sec:statcond}

In the full matrix formalism, in practice we formulate the stationarity condition directly in singular space.
Let $u_{ab;i}$ be the $i$th element of the vector $\v{u}_{ab}$, where $a$, $b$ are the matrix indices just as in Eq.~\eqref{eq:param-B}.
Then, the stationarity condition reads $\partial Q_\alpha/\partial u_{ab;i} = 0$ for all $a$,~$b$,~$i$.
$Q_\alpha$ is given by Eq.~\eqref{eq:matrix-Q}.
Its derivative is
\begin{equation}
 \pd{Q_\alpha}{u_{ab;i}} = \sum_{cd;j}\left[\frac12 \pd{\chi^2 (\v{A}_{cd})}{A_{cd;j}} - \alpha \pd{S(\v{A}_{cd})}{A_{cd;j}} \right] \pd{A_{cd;j}}{u_{ab;i}}.
\label{eq:dQ_du-matrix}
\end{equation}
The derivative of the misfit is
\begin{equation}
    \pd{\chi^2 (\v{A}_{cd})}{ A_{cd;j}}
        = 2 \sum_l \frac{1}{\sigma_{cd;l}^2} \left(\sum_k  K_{lk} A_{cd;k} - G_{cd;l} \right) K_{lj},
\end{equation}
where the data $G_{cd;l}$ are assumed to have diagonal covariance matrices with diagonal elements $\sigma^2_{cd;l}$ (in practice, a change of basis to diagonalize the covariance matrix is always possible).
For the diagonal elements, the derivative of the conventional entropy [Eq.~\eqref{eq:entropy-conventional}] is
\begin{equation}
    \pd{S(\v{A}_{cc})}{A_{cc;j}}
        = - \log \frac{A_{cc;j}}{D_{cc;j}},
\end{equation}
for the off-diagonal elements it is given by Eq.~\eqref{eq:derivative_entropy_pm}.
When using $A_{cd;j} = \sum_e B_{ec;j}^* B_{ed;j}$, one obtains
\begin{equation}
    \frac{\partial A_{cd;j}}{\partial u_{ab;i}}
        = \sum_e \pd{B_{ec;j}^*}{u_{ab;i}} B_{ed;j} + B_{ec;j}^* \pd{B_{ed;j}}{u_{ab;i}}.
\end{equation}
As we know that $\partial B_{cd;j}/\partial u_{ab;i}$ is zero unless $a=c$ and $b=d$, the sum over $e$ drops out and one has
\begin{equation}
    \frac{\partial A_{cd;j}}{\partial u_{ab;i}}
        = \delta_{bc}\pd{B_{ab;j}^*}{u_{ab;i}} B_{ad;j} + \delta_{bd} B_{ac;j}^* \pd{B_{ab;j}}{u_{ab;i}},
\end{equation}
where [using Eq.~\eqref{eq:param-B}]
\begin{equation}
    \frac{\partial B_{ab;j}}{\partial u_{ab;i}}
        = D_{ab;j} V_{ji} \left(e^{\sum_k V_{jk} u_{ab;k}} + e^{-\sum_k V_{jk} u_{ab;k}}\right).
\end{equation}
By plugging these derivatives into Eq.~\eqref{eq:dQ_du-matrix} and setting it to zero, one obtains an expression for the stationarity condition that has to be solved for $u$.

\section{Implementation benchmarks}\label{sec:appl-bench}
\begin{figure}[t]
    \centering
    \includegraphics[width=\linewidth]{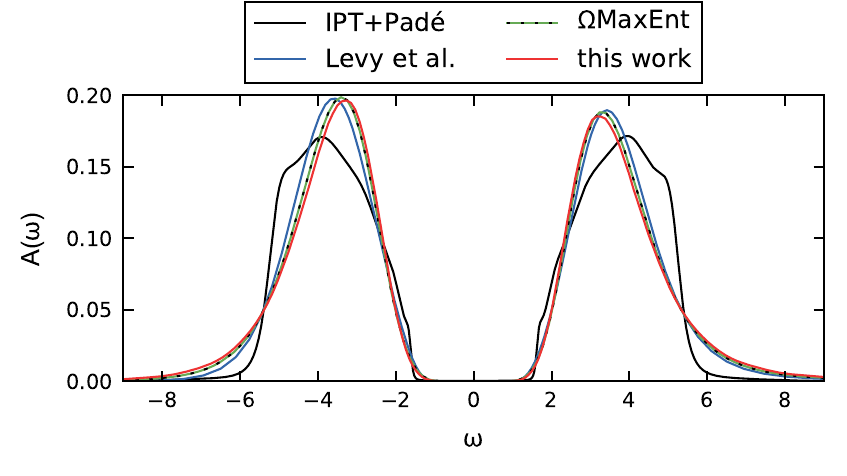}
    \caption{
    Spectral function $A(\omega)$ of the first band of the diagonal two-band model with $U_i/D_i = 3.25$, calculated using Pad\'e approximants for the IPT solution (black) and the MEM for the CTHYB solution.
    Different MEM codes (Bryan solution from Levy \emph{et al.}~\cite{levy_implementation_2016} in blue, $\Omega$\textsc{maxent}~\cite{PhysRevE.94.023303} in dashed green, and our code in red) were used; a flat default model was employed for all three cases.
    }
    \label{fig:model-diag-comp}
\end{figure}
In this appendix, we perform a few tests based on the model introduced in Sec.~\ref{sec:bethe-model} in order to check our implementation of the MEM in general and to demonstrate the effect of noise on the AC.

First, we compare the results for our unrotated diagonal two-band model to those obtained with two freely available MEM codes: a code recently presented by Levy \emph{et al.}~\cite{levy_implementation_2016} and the $\Omega$\textsc{maxent} code~\cite{PhysRevE.94.023303}.
The resulting spectral function for the first band is shown in Fig.~\ref{fig:model-diag-comp}.
Within the errors of the method (as discussed, e.g., in Ref.~\cite{gunnarsson_analytical_2010}), the three MEM curves are in good agreement.
For the second band, the quality of the AC is similar (not shown here).
The fact that the MEM solution and the Pad\'e solution do only qualitatively agree is not surprising, as even a small statistical noise of the Monte Carlo data, in contrast to the noiseless IPT solution, notably increases the uncertainty of the AC~\cite{gunnarsson_analytical_2010}.

\begin{figure}[t]
    \centering
    \includegraphics[width=\linewidth]{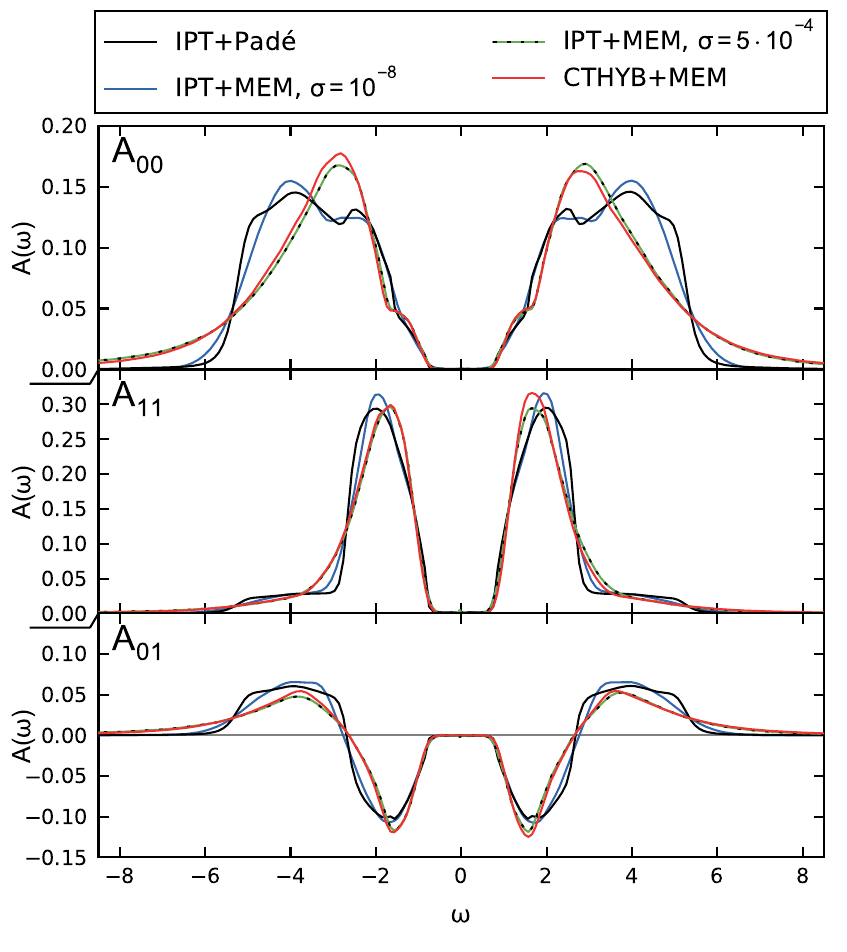}
    \caption{
        Spectral function $A(\omega)$ of the two-band model with $U_i/D_i = 3.25$, calculated using Pad\'e approximants for the IPT solution.
        Starting from that solution, a $G(\tau)$ was calculated using Eq.~\eqref{eq:Gtau-from-Aw-matrix}; that $G(\tau)$ was then rotated in accordance to Sec.~\ref{sec:bethe-model}.
        The rotated IPT and Pad\'e result is shown as the black curve.
        Gaussian random noise with a standard deviation of $\sigma$ was added to that $G(\tau)$, which was then analytically continued using the full matrix MEM (blue curve with $\sigma = 10^{-8}$, dashed green curve for $\sigma=5\cdot 10^{-4}$).
        This is compared to the result from analytically continuing the $G(\tau)$ obtained by solving the same model with CTHYB (red).
        }
    \label{fig:model-backforth}
\end{figure}

In order to assess this influence of statistical noise on the AC, we take the spectral function $A(\omega)$ obtained from IPT and Pad\'e as starting point; then, we calculate the corresponding imaginary-time Green's function $G(\tau)$ by multiplying with the kernel, as in Eq.~\eqref{eq:Gtau-from-Aw-matrix}.
In analogy to Sec.~\ref{sec:bethe-model} we rotate the $G(\tau)$ so that it features off-diagonal elements (the rotated IPT and Pad\'e curve is the black curve in Fig.~\ref{fig:model-backforth}).
From that $G(\tau)$ (a very small Gaussian random noise with standard deviation $10^{-8}$ has to be added for the MEM to work) an $A(\omega)$ can be obtained once more using AC, which we perform using our full matrix MEM (blue curve in Fig.~\ref{fig:model-backforth}).
As can be clearly seen, for all matrix elements the original curve is well reproduced by the MEM, which is further evidence that our implementation works.
However, the curves are smoother than the original data for larger $|\omega|$; this is a well-known tendency of all MEM as the entropy term favors the smoothness of the default model and $G(\tau)$ generally represents the spectral features worse for higher $|\omega|$.

We now emulate QMC data by adding bigger random Gaussian noise (with standard deviation $5\times 10^{-4}$) to the $G(\tau)$ from the IPT and Pad\'e $A(\omega)$.
The MEM-analytically continued curve from that noisy $G(\tau)$ (dashed green curve in Fig.~\ref{fig:model-backforth}) differs considerably both from the input $A(\omega)$ and from the MEM curve with hardly any noise. One can clearly see how much information is lost already by noise only of the order of $5\times 10^{-4}$.
The detailed structure of the Hubbard bands cannot be resolved and it is replaced by just one broad peak for each band (in the diagonal elements).
In the $A_{00}$ element, a small shoulder for low $|\omega|$ is observed, reminiscent of a similar feature in the original data.

Finally, we want to compare the results from our artificially noisy Green's function to that of our CTHYB calculation for the same model.
The spectral functions (dashed green and red curves in Fig.~\ref{fig:model-backforth}) look very similar, especially the off-diagonal element.
As noted before, the CTHYB solution breaks particle-hole symmetry, which can be nicely seen in the different peak heights for positive and negative $\omega$ in the diagonal elements (of course, the particular way how this breaking happens will be different from QMC run to QMC run due to the stochastic nature of the method).
This can already be seen on the level of $G(\tau)$ (not shown).
Thus, we conclude that IPT and Pad\'e gives results compatible to
CTHYB and MEM, for this very specific model at hand.

\vfill

\bibliography{literatur.bib}

\end{document}